\documentstyle[twoside,fleqn,espcrc2,epsfig]{article}

\newcommand{\ktev}{KTeV}
\newcommand{\kz}{\mbox{$K^0$}}
\newcommand{\kzb}{\mbox{$\overline{K^0}$}}
\newcommand{\kl}{\mbox{$K^0_L$}}
\newcommand{\ks}{\mbox{$K^0_S$}}

\newcommand{\pim}{\mbox{$\pi^-$}}
\newcommand{\pip}{\mbox{$\pi^+$}}
\newcommand{\piz}{\mbox{$\pi^0$}}

\newcommand{\eneg}{\mbox{$e^-$}}
\newcommand{\epos}{\mbox{$e^+$}}

\newcommand{\reps}{\mbox{$\rm{Re}(\epsilon'/\epsilon)$}}

\newcommand{\AmS}{{\protect\the\textfont2
  A\kern-.1667em\lower.5ex\hbox{M}\kern-.125emS}}

\title{Status of the \ktev\ Experiment at Fermilab}

\author{R. Ben-David\address{Fermi National Accelerator Laboratory,\\
P.O.Box 500, Batavia, IL, USA, 60510}\thanks{Representing the \ktev\
collaboration.}}
       
\begin{document}

\begin{abstract}
The \ktev\ experiment is a fixed target experiment at Fermilab.  Its
primary goal is the search for direct CP violation in the decay of
neutral kaons.   Its current status and some preliminary results will
be discussed.
\end{abstract}

\maketitle

\section{Introduction}
\label{sec:intro}

The primary goal of the \ktev\ experiment is the search for direct CP
violation in the neutral kaon system.  It is believed that CP is
violated through two different processes.  The first process 
has been observed and has been termed indirect CP violation.  It
 is the result
of mixing in the mass matrix through $\kz-\kzb$ mixing.  The second
process,  called direct CP violation, occurs in the decay amplitude.

The \ktev\ experiment is a fixed target experiment at Fermilab.  It
is, in fact, two different experiments.  The first is
E799-II, which searches for CP violation  through the  rare
decay processes $\kl \rightarrow \piz l \overline{l} \; (l = e, \mu,
\nu)$. These processes are expected to have a large CP violating
amplitude, in contrast to $\kl \rightarrow \pi\pi$, where the direct
CP violating amplitude  is $\ll 10^{-2}$ of the indirect CP
violating amplitude.
Table~\ref{tab:rare_decays} show's the current theoretical
\begin{table*}[hbt]
\setlength{\tabcolsep}{0.8pc}
\newlength{\digitwidth} \settowidth{\digitwidth}{\rm 0}
\catcode`?=\active \def?{\kern\digitwidth}
\caption{Branching Ratios for the direct CP violating modes
$\kl \rightarrow \piz l \overline{l}$. }
\label{tab:rare_decays}
\begin{tabular*}{\textwidth}{@{}l@{\extracolsep{\fill}}rrrr}
\hline
		   \multicolumn{1}{c}{Decay mode}
                 & \multicolumn{1}{c}{Theory} 
                 & \multicolumn{1}{c}{Experimental limit} 
                 & \multicolumn{1}{c}{E799-II's goal in 1997}\\
\hline 
 $\kl \rightarrow \piz \nu \overline{\nu}$  
              & $ (2.8\pm1.7)\times10^{-11}$~\cite{buchalla} 
              & $ <5.8\times10^{-5}$~\cite{weaver}
              & $ <1.8\times10^{-7}$ \\
 $\kl \rightarrow \piz e^+ e^-$  
              & $ 10^{-11}-10^{-12}$~\cite{dono} 
              & $ <4.3\times10^{-9}$~\cite{harris1}
              & $ <2.5\times10^{-10}$ \\
 $\kl \rightarrow \piz \mu^+ \mu^-$  
              & $ 6.3\times10^{-12}$~\cite{ecker} 
              & $ <5.1\times10^{-9}$~\cite{harris2} 
              & $ <1.6\times10^{-10}$ \\

\hline
\multicolumn{4}{@{}p{120mm}}{The upper limits are presented at the
90\% C.L.}
\end{tabular*}
\end{table*}
predictions and experimental upper limits on the branching ratios,
along with the expected sensitivity for E799-II.
In addition, E799-II will also search for and/or
measure other rare or forbidden decays.  The other experiment is E832,
which 
endeavors to measure the direct component of $\kl \rightarrow \pi\pi$,
parameterized as $\epsilon'$.

E832 measures the double ratio of the four $K_{L,S}\rightarrow \pi\pi$
decay rates to determine \reps\ through the expression:
\begin{eqnarray}
\lefteqn{\frac{\Gamma(\kl\rightarrow\pip\pim)/\Gamma(\ks\rightarrow\pip\pim)}{\Gamma(\kl\rightarrow\piz\piz)/\Gamma(\ks\rightarrow\piz\piz)}
 \approx }  \hspace{1.5in}    \nonumber \\
&   &  1 + 6\reps.
\end{eqnarray}
Current experimental measurements of \reps\ are:
\[ \reps = \left\{ \begin{array}{ll}
 (23\pm3.5\pm6.0)\times 10^{-4} & \rm{NA31}~\cite{na31} \\
 (7.4\pm5.2\pm2.9)\times 10^{-4} & \rm{E731}~\cite{e731},
\end{array}
\right. \]
while the theoretical predictions cover the range $(-50 \rightarrow
+43)\times 10^{-4}$~\cite{buras}.
  \ktev's goal is to measure \reps\ to a precision
of $\sim\!1\times 10^{-4}$.


In September of 1997, the Fermilab fixed target program completed a 
successful 15 month run.  During that run, each of the two experiments
had two running 
periods.  E832 ran first, for seven weeks,
accumulating roughly four times the statistical sample of the previous
generation of the experiment, E731.
E799-II ran next for eight weeks, almost quadrupling the number of
kaons observed in E799-I.  
Taking both E832 running periods into account, approximately 16
times the  statistical sample of E731 was accumulated.
  E799-II had a slightly shorter
second run of 6 weeks, and almost doubled the statistical sample
gathered in its  first run.  Most of what will be discussed in these
proceedings will be based on data collected during the first running period
of each experiment.

\section{The KTeV Detector}

\begin{figure*}[htb]
\vspace{9pt}
\epsfig{file=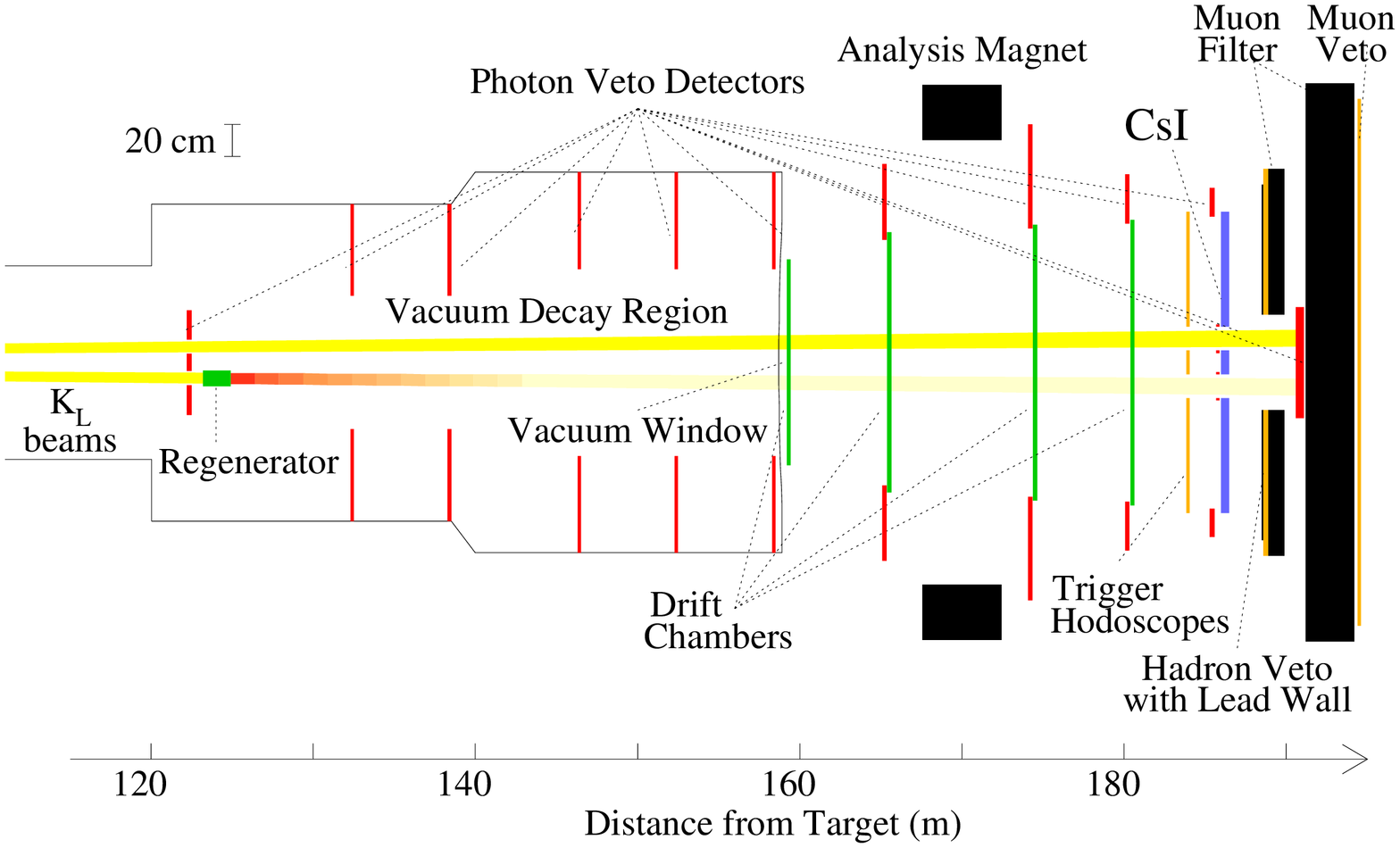, height=\linewidth, width=3in,angle=270.0,bbllx=50,bblly=50,bburx=450,bbury=750}
\caption{The \ktev\ spectrometer.}
\label{fig:detector}
\end{figure*}

To reduce detector performance dependence  effects, E832 measures all four
$K_{L,S}\rightarrow \pi\pi$ decay rates 
simultaneously.  The charged
mode decays are measured with a charged spectrometer, consisting of
four drift chambers and an analysis magnet (two chambers on each side
of the magnet).  The neutral mode decays are measured with an
electromagnetic calorimeter.  A series of photon veto scintillation
counters
 are
used to detect photons escaping the fiducial volume (see
figure~\ref{fig:detector}).

To simultaneously create both \kl s and \ks s, two parallel neutral
kaon beams are 
created.  The KTeV detector is located almost 30 \ks\ lifetimes
downstream from the 
target so that two parallel \kl\ beams arrive at the detector.
  At 124~meters from the production target, one
of the beams passes through a regenerator producing \ks s.  To
reduce acceptance
 biases, the regenerator moves between the two
beams, once per minute.
 Both the \kl s and
\ks s decay inside the vacuum vessel.

For E799-II, the regenerator and the photon veto detector upstream of
it are removed.  To improve the $\pi/e$ rejection, a necessity for
rare kaon decay measurements, eight planes of
transition radiation detectors are placed directly after the most
downstream drift chamber.  They provide an additional 200:1 $\pi/e$
rejection to the roughly 500:1 rejection of the calorimeter.

\section{E832}

\subsection{Measurement Technique}

To achieve the design precision on the measurement of \reps\ to
$1\times10^{-4}$, both the statistical and systematic errors must be
reduced relative to the E731 measurement.

To reduce the statistical error, several improvements were made to the
experiment: the experiment ran with a more
intense beam, the detector was upgraded to enable running at
higher intensity and trigger rates, the data acquisition live time was 
increased, and the the experiment took data for a longer period
of time.

There are four dominant contributions to the systematic error:
background events, uncertainty in the 
energy scale, acceptance, and accidental activity.
Several detector and beam improvements 
 will reduce background events: a new beamline design that was
significantly cleaner than before, a fully active regenerator that
will better separate events from interactions in the regenerator, a
hermetic photon veto system, and a high resolution CsI calorimeter.
The energy scale of the calorimeter will be calibrated to high
precision using electrons from a large $\kl \rightarrow \pi e \nu$ sample.
 The detector acceptance for charged mode and neutral
mode decays will be studied using the large samples of  $\kl
\rightarrow \pi e \nu$ and  $\kl \rightarrow 3\piz$ decays, respectively.
Accidental activity, that is, activity not associated with the event,
was reduced because of the improved beam design, the fully active
regenerator, and improved timing resolution.
It is expected that the combined systematic error on \reps\ will be
less than $1\times10^{-4}$.
\subsection{Status}

\begin{figure}[htb]
\vspace{9pt}
\epsfig{file=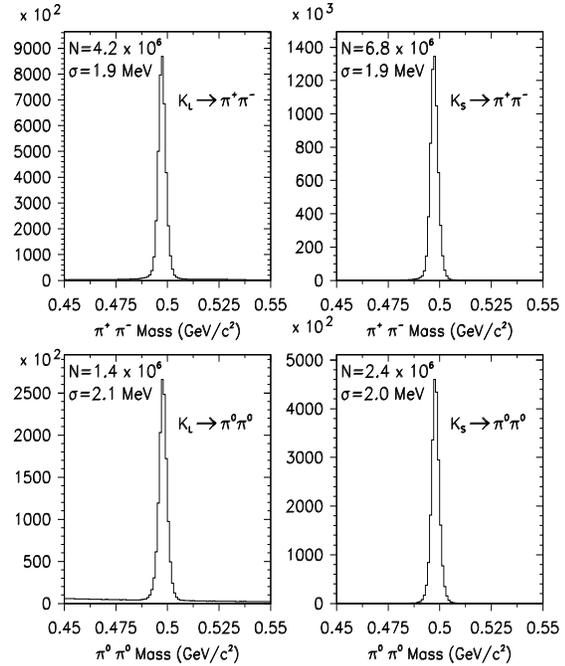, height=3.5in, width=\linewidth,bbllx=70,bblly=170,bburx=545,bbury=670}
\caption{Invariant mass distributions for the four decay modes contributing
to the measurement of \reps.  The decays from the beam going through the
regenerator
are in the  histograms in the right column, while the decays from  the vacuum
beam are in 
the histograms in the left column.}
\label{fig:2pi}
\end{figure}

Figure~\ref{fig:2pi}
shows the online invariant-mass plots for the four $K_{L,S} \rightarrow
\pi\pi$ decays using
statistics  accumulated
over the first running period.  Offline analysis will remove
approximately 30\% of the events and reduce the backgrounds by an
order of magnitude.  A more thorough calibration of the detector will
improve the mass resolution by $\sim\!10\%$.

  Combining both
runs and after applying offline cuts, E832 will have accumulated
 $\sim\!4\times10^{6}$ vacuum \piz\piz\ decays (this is the
mode that statistically limits the measurement).  This will give a
statistical error on \reps\ of $\sim\!1.3\times10^{-4}$.

\section{E799-II}

In addition to the two running periods described in
section~\ref{sec:intro}, E799-II 
 took data for one day, in a special configuration,
optimized to search for $\kl \rightarrow \piz \nu \overline{\nu}$.

\subsection{Search for  $\kl \rightarrow \piz \nu \overline{\nu}$}

The branching ratio of  $\kl \rightarrow \piz \nu \overline{\nu}$ is
directly proportional to the scale of CP violation in the Standard Model.
The current status of the measurement is given in
table~\ref{tab:rare_decays}. 


The signature for the decay is a single \piz\ with large transverse
momentum, $P_T$, relative to the direction of the decaying \kl.  
The 
detector configuration was optimized to search for a single \piz\
during a special one day run.  A
single beam, $4\times4~\rm{cm}^2$ in size was used.  In addition, the
regenerator and transition radiation detectors were removed.

\begin{figure}[htb]
\vspace{9pt}
\epsfig{file=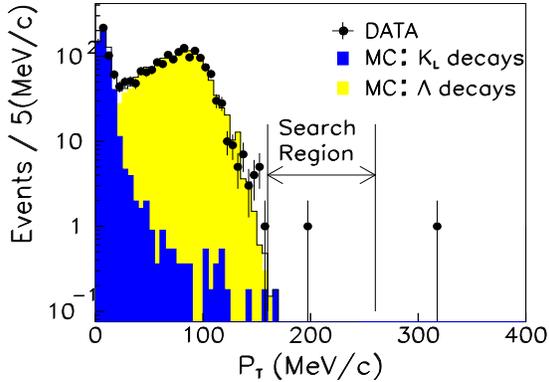, height=2in, width=\linewidth,bbllx=20,bblly=0,bburx=540,bbury=374}
\caption{The \piz\ transverse momentum distribution in the search for
$\kl\rightarrow\piz \nu \overline{\nu}$, after all cuts except for the
$P_T$ cut have been applied.  The data are overlayed on
the simulation of the two dominant background modes.  The events with
a peak at zero come from the $\kl\rightarrow\gamma\gamma$ decay.}
\label{fig:piznunu}
\end{figure}

Figure~\ref{fig:piznunu}
 is the $P_T$ distribution of the reconstructed \piz s.
The signal region requires the \piz\ to have a  $P_T$ between
160 and 260~MeV/c, in order
 to be above the kinematic limit of \piz s from the decay
$\Lambda \rightarrow \piz n$ and within the upper kinematic limit of the
decay.
There are two events with  $P_T$ greater than 160~MeV/c.
One is in the signal region and the other is above it.
  Preliminary analysis of the data shows that the source of
the events is most likely 
an interaction of a beam neutron with the detector, $n +
n(p) \rightarrow n + 2\piz + X$, where a photon from each \piz\ is
lost and the remaining two photons form a \piz\ at the wrong $z$
position and large $P_T$.
 
The analysis has not yet been completed, however, there is a preliminary
result for the branching ratio.
  Using the decay $\kl \rightarrow \piz \piz$ as the
normalization mode, the single event sensitivity for $\kl \rightarrow \piz \nu
\overline{\nu}$ is $(4.4\pm0.1)\times10^{-7}$, where the event in the 
signal region is treated as  background and the error quoted is
entirely due to statistics.
This gives an upper limit on the branching ratio of
$1.8\times10^{-6}$ at the 90\% confidence level.  This result is
$\sim\!30$ times more sensitive than the previous result.

\subsection{$\kl \rightarrow \pip\pim\epos\eneg$}

In addition to searches for direct CP violation at KTeV, there are
 searches
 for rare kaon decay modes.  One such decay mode is $\kl
\rightarrow \pip\pim\gamma^* \rightarrow \pip\pim\epos\eneg$.  The
interest in this mode arises from the fact that the emitted $\gamma^*$s
have a net circular polarization.  
This polarization arises from the interference of  CP conserving and
 CP violating amplitudes.
One amplitude is due to the inner bremsstrahlung of the CP violating
 $\kl
\rightarrow \pip\pim$ decay.  The other
amplitude is from the CP conserving direct emission process.  In
the final state, the two amplitudes interfere with each other, causing
 the $\gamma^*$ to be emitted with a net circular polarization.  
The net polarization can be determined by
 measuring an asymmetry in the angle between
 the electron and pion planes in the center of
mass.   The asymmetry
is expected to be sensitive to indirect CP violation, however
  any deviation
from the theoretical expectation could be a sign of new physics.
The most recent experimental upper limit on the branching ratio is
$4.6\times10^{-7}$, at the 90\% confidence level~\cite{nomura}, while the
theoretical prediction is $\sim\!3\times10^{-7}$~\cite{sehgal}.

There are several signatures to the 
decay $\kl \rightarrow \pip\pim\epos\eneg$: 
four tracks originating at a common vertex,
opposite signed $\pi$s and $e$s,
the resultant momentum vector of the decay products should have a
small component relative to a line connecting the target and the decay
vertex, and the event must pass a kinematic cut designed to remove the
dominant background mode $\kl \rightarrow \pip\pim\piz \rightarrow
\pip\pim\epos\eneg\gamma$.

\begin{figure}[htb]
\vspace{9pt}
\epsfig{file=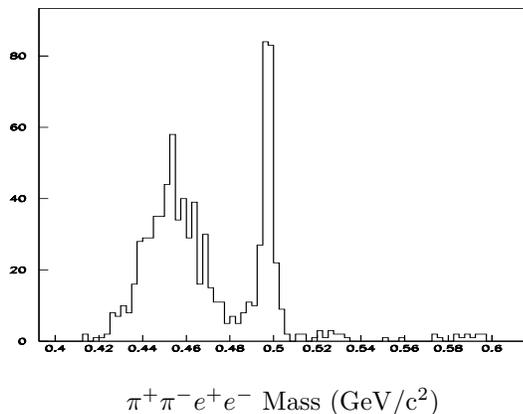, height=2in, width=\linewidth,bbllx=40,bblly=150,bburx=540,bbury=660}
\hspace*{.75in}$\pip\pim\epos\eneg$ Mass ($\rm{GeV/c^2}$)
\caption{The reconstructed $\pip\pim\epos\eneg$ invariant mass for
approximately one sixth of the total data sample collected.}
\label{fig:pipiee}
\end{figure}

Figure~\ref{fig:pipiee} 
shows the $\pip\pim\epos\eneg$ invariant mass
distribution for approximately one sixth of the full E799-II data
set.  The  $\kl \rightarrow \pip\pim\epos\eneg$ signal 
is clearly visible at the
kaon mass.  The events at lower invariant mass are from the
background mode $\kl \rightarrow \pip\pim\eneg\epos\gamma$ decay,
where the photon was missed.

A preliminary analysis of one day's worth of data yields $36.1\pm6.4$
signal events.  This determines  the branching ratio to be
$(2.6\pm0.5)\times10^{-7}$, where the statistical and systematic errors
have been combined.

\section{Summary}

The KTeV experiment collected a very high quality data set at the
design beam intensity.  Over the next few years, the Fermilab kaon
program will produce a multitude of significant results.  Although the
data collected thus far for E832 will not be enough to reach the
design sensitivity on \reps, another run is scheduled for 1999. It is
expected that the data sample collected
during this past run, will be doubled, thus lowering the statistical error on
\reps\ to $0.9\times10^{-4}$.

The rare kaon program has also collected a large data set.  Some of
the decays modes under study  have 100 times more events than 
previous measurements, so that definitive
measurements of many rare processes will be made.

\end{document}